\documentclass[conference]{IEEEtran}

 \usepackage{lineno,hyperref}
\usepackage{setspace}
\usepackage{amsmath}
\usepackage{graphicx}
\usepackage{algorithm}
\usepackage{geometry}
\usepackage{color}
\usepackage{amsmath}
\usepackage{amssymb}
\usepackage{amsfonts}
\usepackage{caption}
\usepackage{subcaption}
\usepackage{algpseudocode}
\usepackage{amsmath}
\usepackage{cases}
\usepackage{multirow}
\usepackage{etoolbox}
\usepackage{graphicx}
 \usepackage{epstopdf}
\hypersetup{draft}

\usepackage{stfloats}
  \modulolinenumbers[5]

\BeforeBeginEnvironment{lemma}{\noindent}

\usepackage{epsfig}
\graphicspath{{./plots/}, {./figures/}}

\IEEEoverridecommandlockouts
\vspace{-0.3in}
\title{ On Energy Harvesting of Hybrid TDMA-NOMA  Systems
}

\author{\IEEEauthorblockN{  Haitham Al-Obiedollah\IEEEauthorrefmark{1},   Kanapathippillai Cumanan\IEEEauthorrefmark{1}, Alister G. Burr\IEEEauthorrefmark{1},  Jie Tang\IEEEauthorrefmark{2}, \\  Yogachandran Rahulamathavan\IEEEauthorrefmark{3}, Zhiguo Ding\IEEEauthorrefmark{4}, and  Octavia A. Dobre\IEEEauthorrefmark{5} } 
\IEEEauthorblockA{\IEEEauthorrefmark{1}Department of Electronic Engineering, University of York, York, YO10 5DD, UK 
\IEEEauthorblockA{\IEEEauthorrefmark{2} School of Electronic and Information Engineering, South China University of Technology, Guangzhou, China} 
\IEEEauthorblockA{\IEEEauthorrefmark{3} Institute for Digital Technologies, Loughborough University, London E15 2GZ, U.K.} 
\IEEEauthorblockA{\IEEEauthorrefmark{4} School of Electrical and Electronic Engineering, The University of Manchester,
Manchester, UK} 
\IEEEauthorblockA{\IEEEauthorrefmark{5}Department of Electrical and Computer Engineering, Memorial University, St. John’s,  Canada} 
Email: \{hma534, kanapathippillai.cumanan, alister.burr\}@york.ac.uk\IEEEauthorrefmark{1},   eejtang@scut.edu.cn\IEEEauthorrefmark{2},\\ Y.Rahulamathavan@lboro.ac.uk\IEEEauthorrefmark{3},   zhiguo.ding@manchester.ac.uk\IEEEauthorrefmark{4}, odobre@mun.ca\IEEEauthorrefmark{5}}
}

\begin{document}
\maketitle
\linespread{.910}
 
\begin{abstract}
In this paper, we investigate   energy  harvesting   capabilities of   non-orthogonal multiple access (NOMA) scheme integrated with the conventional time division multiple access (TDMA) scheme, which is referred to as hybrid TDMA-NOMA system. In a such  hybrid scheme,   users are divided into a number of  groups,  with the total time allocated for transmission is shared between these groups through multiple time slots. In particular,   a time slot  is assigned to serve each group, whereas the users in the corresponding group are served based on power-domain NOMA technique. Furthermore,         simultaneous wireless power and information transfer   technique is utilized to simultaneously harvest   energy and decode   information at each user. Therefore, each  user splits the received signal into two parts, namely, energy harvesting   part and information decoding   part. In particular, 
  we jointly  determine the power allocation  and   power splitting ratios for all  users to minimize the transmit  power  under   minimum rate  and minimum   energy harvesting requirements at each user.   Furthermore, this joint design  is a non-convex problem in nature. Hence, we employ successive interference cancellation to overcome these non-convexity issues and determine the design parameters (i.e., the power allocations and the power splitting ratios).  In simulation results, we demonstrate  the performance   of   the proposed hybrid TDMA-NOMA design and show that it outperforms  the conventional TDMA scheme in terms of  transmit power consumption. 
\end{abstract}
 
 
\section{Introduction}
In recent years, non-orthogonal multiple access (NOMA) has been envisioned
 as a promising multiple access candidate towards the fifth generation (5G) and beyond wireless networks \cite{fnoma}. In contrast to 
 the conventional orthogonal multiple access (OMA) schemes; namely   time division multiple access (TDMA) and orthogonal frequency division multiple access (OFDMA),  NOMA has the potential   to serve multiple users simultaneously within the same resource block (RB) \cite{fayzeh} \cite{cuma8}. In particular, this can be accomplished  by employing superposition coding   at the base station (BS) \cite{noma99}, such that the messages intended to different users are encoded with different power levels. Furthermore, successive interference cancellation (SIC) is exploited at receiver ends for multiuser detection \cite{cuma} \cite{noma99} \cite{cuma2}. 
 Considering the fact that NOMA can serve multiple users   simultaneously in the same RB,  NOMA is expected to play a crucial role to support the proliferation of Internet-of-Things (IoT) in future wireless networks  \cite{noma3}. However,  the error propagation introduced in  the  SIC process might impose some constraints on the practical implementation of NOMA especially in   dense networks \cite{noma5g}.
  To mitigate this  error propagation issue  and to  facilitate practical implementation of NOMA in dense network, NOMA can be  incorporated with different key technologies to meet various system requirements \cite{islam2017power}. For example, NOMA is integrated with spatial domain multiple access (SDMA); in such a hybrid SDMA-NOMA approach \cite{sdma}, multiple users can be grouped in a cluster and served based on NOMA, while    a spatial beam is assigned to establish communication with  each cluster. Furthermore, NOMA  can be applied  with the existing   conventional OMA schemes, including  hybrid TDMA-NOMA  and hybrid OFDMA-NOMA \cite{noma5g}. In such hybrid schemes, an orthogonal  RB (i.e., time slot or frequency slot) is assigned to serve a group of users, whereas the users in each group are served based on NOMA. These hybrid    systems can offer different advantages over the conventional systems. Firstly, 
 the integration of NOMA with the SDMA/OMA schemes provides additional  degrees of freedom which can be  exploited with the  available domains \cite{noma5g}, \cite{sdma}. 
Secondly, it is obvious that users' grouping (i.e., clustering) minimizes the effects of inevitable errors associated with  SIC by significantly reducing the required number of SIC implementations \cite{zeng2018energy}. \\
 \indent However, massive connectivity   offered by these hybrid schemes has a direct negative impact on the power consumption of these systems. As a result, an  explosive growth in the power consumption is inevitable \cite{green}, which  brings up different environmental issues including global warming and natural disasters, as well as financial pressures on both service providers and consumers \cite{green}. To address these issues, the power consumption in wireless transmission   is mainly considered  by either allocating the power resources to maximize the energy efficiency of the systems  \cite{green} \cite{ee1} \cite{haitham}, or by incorporating the novel   simultaneous wireless power and information transfer (SWIPT) technique \cite{swipt0}. In  SWIPT, the receiver  has the capability  to simultaneously harvest  energy and decode information \cite{swipt0}. In particular, this could be accomplished by splitting the received radio frequency (RF) signal through either time splitting    or  power splitting   techniques \cite{swipt2}. Despite of the  low complexity of the former, this requires a better synchronization between the  receiver and the transmitter to precisely perform the splitting  \cite{swipt0}, and thus, the latter is typically more desirable.  In fact, SWIPT is expected  to contribute in feeding the power-hungry  users, especially  in ultra-dense sensor networks, where hundreds of unreachable sensors seek  power to extend their life-time \cite{swipt0}.  In particular, the co-channel interference introduced  by exploiting the same RB to serve multiple users in the hybrid TDMA-NOMA has the potential to  improve the energy harvesting (EH) capability  compared to the conventional TDMA systems.\\
 \indent In this paper, we investigate the  EH  capabilities of the multi-user single-input single-output (SISO) hybrid TDMA-NOMA system. In particular, we   first introduce the hybrid TDMA-NOMA system model considering the EH capability of each user.   Then, we evaluate the required minimum transmit power  at   BS to meet  the quality-of-service (Qos) requirements. In fact, these requirements include the minimum rate   and minimum harvested power   at each user. The design parameters, i.e., the power allocations and the power splitting ratios, corresponding to the minimum transmit power  are jointly determined  through solving a power minimization (P-Min) problem for the hybrid TDMA-NOMA scheme. However, due to the non-convexity of  the P-Min problem, sequential convex approximation (SCA) is exploited  to  \textit{jointly} optimize these design parameters. 
 Finally, we compare the performance of    the proposed hybrid NOMA-TDMA design   with that of the conventional TDMA scheme.\\
\indent The remainder of the paper is organized as follows.  Section \ref{sec2}  introduces  the system model and   the problem formulation. Section \ref{sec3} presents  the proposed technique to solve the formulated
   optimization problem,  with additional discussions on the TDMA-NOMA system.  Section \ref{sec4}  provides simulation results which evaluate the effectiveness of the proposed joint design by comparing its performance with that of the conventional  TDMA scheme. Finally,   Section\ref{sec5}   concludes the paper.

 \section{System Model and Problem Formulation}\label{sec2}
 \subsection{System Model}
 
\begin{figure}[H]
\centering
\includegraphics[width=.9  \linewidth]{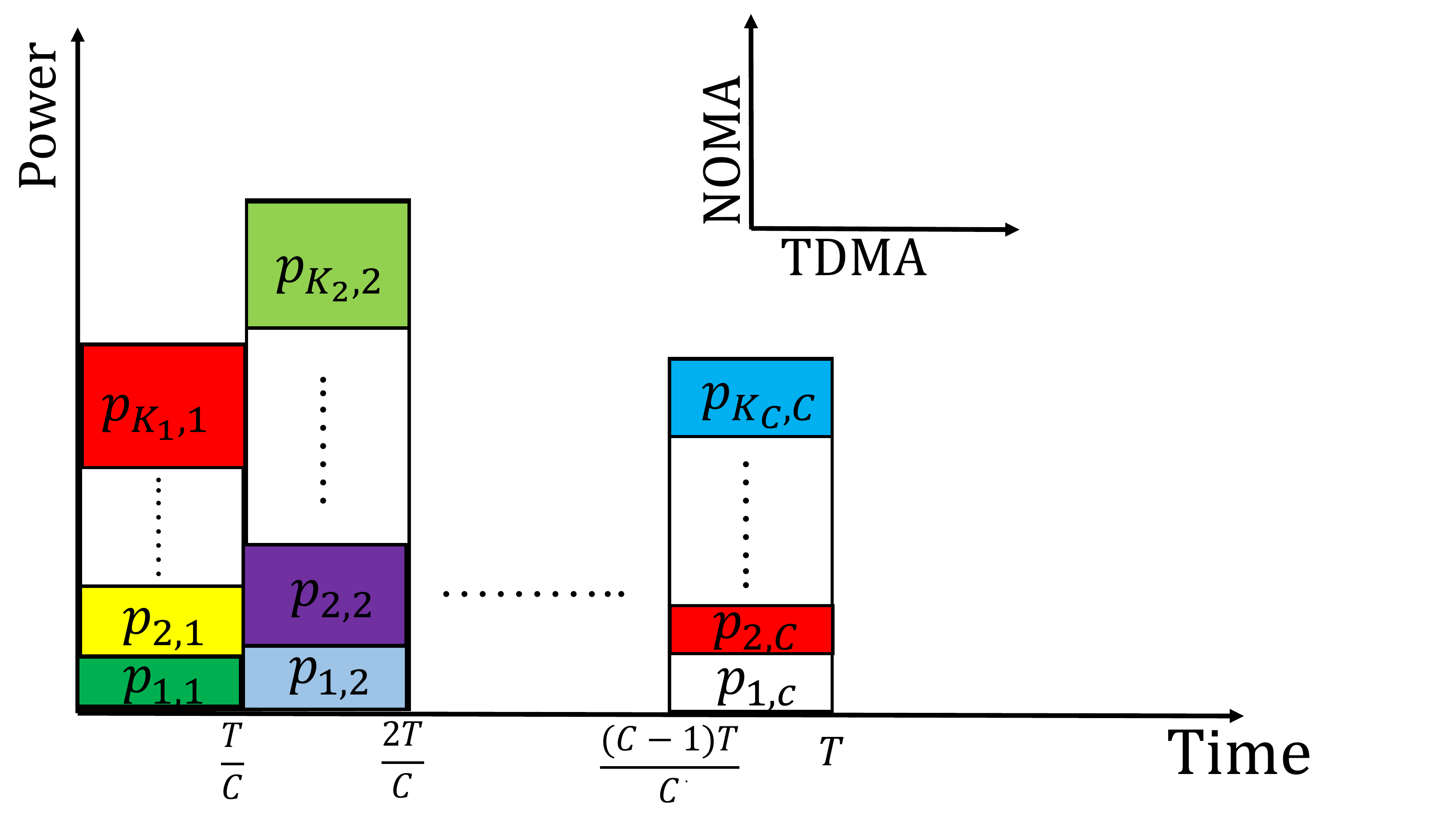}
\caption{Hybrid TDMA-NOMA system; the users are divided among   $C$ groups.}
\label{tdma-noma}
\end{figure}

We   consider a multi-user SISO hybrid TDMA-NOMA    system with $K$ users, where the $k^{th}$ single-antenna user ($u_k$) is located at a distance $d_k$ (in meter) from the BS. Furthermore, the users are divided into $C$ groups, and the available time for transmission ($T$) is divided equally among these groups, as shown in Fig.~\ref{tdma-noma}.  The number of   users in the $i^{th}$ group ($G_i$)  is denoted by $K_i$, $\forall i\in \mathcal{C} \overset{\Delta}{=}\{1,2,\cdots,C\}$, such that $K=\sum_{i=1}^{C} K_i$. Note that the time slot allocated to serve group $G_i$ is denoted by  $t_i$, such that $T=\sum_{i=1}^{C} t_i$, and $t_i=\frac{T}{C}$.   Furthermore, the  $j^{th}$ user in  $G_i$ is denoted by  $u_{j,i}$.   
In particular, user grouping has a considerable impact on the performance of the hybrid TDMA-NOMA system. Hence, we discuss the user grouping strategies in the next section. 
\setlength{\belowcaptionskip}{-10pt}

Furthermore,    
multiple users in each group are served based on NOMA, such that  the transmit signal at the BS at $t_i$ is denoted by     $x_i$, expressed as 
\begin{equation}\label{tran1}
x_{i}=\sum_{j=1}^{K_i} p_{j,i}x_{j,i}.
\end{equation}
 Note that  $p_{j,i}^2$ and $x_{j,i}$ denote the  allocated power and the signal intended to the $j^{th}$ user in $G_i$ (i.e., $u_{j,i}$), respectively. Based on this, the received signal at $u_{j,i}$ can be expressed as  
\begin{equation}\label{rec}
 r_{j,i}=h_{j,i}x_i+n_{j,i},  \forall i\in \mathcal{C}, \forall j\in \mathcal{K}_i\overset{\Delta}{=} \{1,2,\cdots, K_{i}\},
 \end{equation}
 where $h_{j,i}$ is the   channel coefficient between the BS and $u_{j,i}$. In particular,  the corresponding  channel gain  can be written as $|h_{j,i}|^2= \frac{\eta}{({ {d_{j,i}}/{d_o})}^{\kappa}}$ \cite{swipt10}, where $d_o$ and  $d_{j,i}$  represent the reference distance and the  distance between $u_{j,i}$ and the BS (in meters),  respectively. Furthermore, $\eta$ and $\kappa$ represent the signal attenuation at     $d_o$ and the path loss exponent, respectively. In addition,  $n_{j,i}$ is the additive white Gaussian noise (AWGN) with zero mean and variance $\sigma_{j,i}^2$ dBm/Hz.     
 In fact, user  ordering at each group plays a crucial role on the performance of NOMA systems \cite{islam2018}. Note that  the optimal ordering can be determined through an exhaustive search among all   ordering possibilities, which is not possible in practical scenarios, especially  in dense networks \cite{wcnc1}, \cite{wcnc2}. Hence,     we order the users in  each group  based on their channel strengths \cite{mimonoma}, \cite{haitham}, as follows:
\begin{equation}\label{order}
|h_{1,i}|^2 \geq|h_{2,i}|^2 \geq\cdots\geq|h_{K_i,i}|^2, \quad  \forall i \in \mathcal{C}.
\end{equation} 
\setlength{\belowcaptionskip}{-10pt}

\begin{figure}[h]
\centering
\includegraphics[width=.70\linewidth]{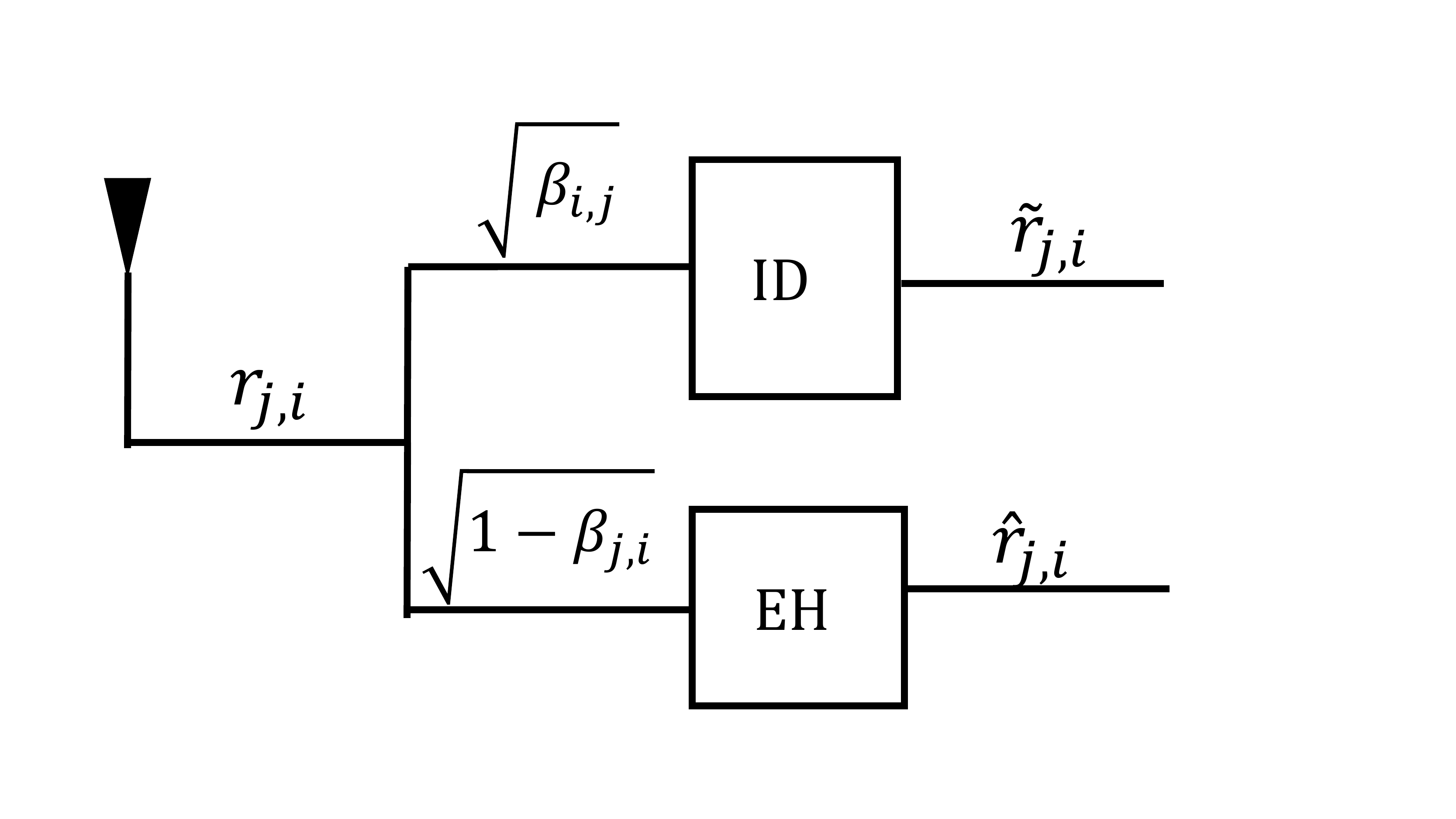}
\caption{Illustration of splitting the received signal at $u_{j,i}$.}
\label{fig4}
\end{figure} 
Now, we assume that each user has a potential capability to split the received signal into two parts such that  $\sqrt{\beta_{j,i}}$ (i.e., $0<\beta_{j,i}<1$) of $r_{j,i}$ is used to decode information, whereas $\sqrt{1-\beta_{j,i}}$ of $r_{j,i}$ is utilized to harvest energy,  as shown in Fig.~\ref{fig4}.

\subsubsection*{\textbf{Information Decoding (ID) Stage}}
At this stage, the fraction of received signal  $\sqrt{\beta_{j,i}}$ of $r_{j,i}$ is utilized to decode the information. Therefore,  the  { signal split to} the ID stage can be written as follows:
 \begin{multline}
\tilde{r}_{j,i}= \sqrt{\beta_{j,i}} \left(h_{j,i}\sum_{s= 1}^{K_i} p_{s,i}x_{s,i}+n_{j,i}\right)+\tilde{n}_{j,i},  \\ \forall i \in \mathcal{C}, \forall j \in \mathcal{K}_i,
 \end{multline}
 where $\tilde{n}_{j,i}$ is  AWGN with zero mean and variance $\tilde{\sigma}_{j,i}^2$ dBm/Hz,  which is introduced due to processing $ r_{j,i}$ on the ID stage \cite{swipt4}, \cite{swipt0}, \cite{ma2019generic}. Based on  the  user ordering defined in  (\ref{order}),  $u_{j,i}$ performs SIC to detect and remove the signals of the weaker users $u_{j+1,i} \cdots u_{K_i,i}$ prior to decoding its own signal. Note that we assume that SIC is  implemented with no errors.  Hence, the received signal at  $u_{j,i}$  after employing SIC can be written as follows:
 \begin{multline}
\tilde{r}_{j,i}^{\text{SIC}}=\sqrt{\beta_{j,i} }\left(h_{j,i}p_{j,i}x_{j,i}+h_{j,i}\sum_{s= 1}^{j-1} p_{s,i}x_{s,i}+n_{j,i}\right)\\+\tilde{n}_{j,i}, \forall i \in \mathcal{C}, \forall j \in \mathcal{K}_i.
 \end{multline}
Accordingly,   the signal-to-interference and noise ratio (SINR) at which  $u_{j,i}$    decodes the message of $u_{d,i}$ $\forall d \in \{j+1,\cdots,K_i\}$ can be written as 
 \begin{multline}
 \text{SINR}_{j,i}^{d}=\frac{\beta_{j,i}|h_{j,i}|^2p_{d,i}^2}{\beta_{j,i} |h_{j,i}|^2\sum_{s= 1}^{d-1} p_{s,i}^2+\beta_{j,i} \sigma_{j,i}^2+\tilde{\sigma}_{j,i}^2 }\\ \forall i \in \mathcal{C}, \forall j \in \mathcal{K}_i, \forall d \in \{j+1,j+2,\cdots, \mathcal{K}_i\}.
 \end{multline}  
Note that $u_{j,i}$ has the capability to decode the message of the weaker user $u_{d,i}$ $\forall d \in \{j+1,\cdots,K_i\}$  if and only if the messages intended to these weaker users are received at $u_{j,i}$ with higher SINR { compared to  that of the users with stronger channel conditions}. In particular, this can be guaranteed by including the following condition:
\begin{equation}\label{sic} 
 p_{K_i,i}^2\geq p_{K_i-1,i}^2\geq\cdots\geq p_{1,i}^2, \forall i \in \mathcal{C}.
\end{equation}
The above constraint   is referred to as SIC constraint throughout this paper. 
  Based on this condition, the SINR of $u_{j,i}$ can be defined as  \cite{hanif}
 \begin{multline}
 \text{SINR}_{j,i}=\text{min}\{\text{SINR}_{j,i}^1,\text{SINR}_{j,i}^2,\cdots,\text{SINR}_{j,i}^j \},\\ \forall i \in \mathcal{C}, \forall j \in \mathcal{K}_i.
 \end{multline}
 Therefore, the achieved rate at $u_{j,i}$ can be written as follows:
 \begin{equation}
 {R}_{j,i}= t_i \log_2(1+\text{SINR}_{j,i}), \forall i \in \mathcal{C}, \forall j \in \mathcal{K}_i.
 \end{equation}
  Note that the total required transmit power   at the BS can be expressed as $P_t= \sum_{i=1}^{C}  \sum_{j=1}^{K_i} p_{j,i}^2$.
 \subsubsection*{\textbf{EH Stage}}
 Next, we consider the received signal part at the EH stage. { In particular, the EH circuit consists of a matching network, a radio frequency to direct current (RF-DC) converter, and a storage unit \cite{newref} \cite{swipt1}}.  The signal split to the EH stage can be written as
 \begin{multline}
  \hat{r}_{j,i}=\sqrt{\left(1-\beta_{j,i}\right)} \left(h_{j,i}\sum_{s= 1}^{K_i} p_{s,i}x_{s,i}+ { {n}_{j,i}}\right)+ \hat{n}_{j,i},\\ \forall i \in \mathcal{C}, \forall j \in \mathcal{K}_i,
 \end{multline}
 where $\hat{n}_{j,i} $ is AWGN  introduced by the processing  of the received signal at the EH stage with zero mean and variance  $\hat{\sigma}_{j,i}^2$ dBm/Hz. Note that $\hat{r}_{j,i}$ is utilized to harvest the energy at $u_{j,i}$.
  Hence, the harvested power at $u_{j,i}$  can be defined as \cite{swipt2}
 \begin{equation}
P_{j,i}=\eta \left(1-\beta_{j,i}\right)  \left(|h_{j,i}|^2\sum_{s= 1}^{K_i} p_{s,i}^2\right),  \forall i \in \mathcal{C}, \forall j \in \mathcal{K}_i,
 \end{equation}
 where $\eta  \in [0,1]$ is the  efficiency of the RF-DC   in the EH stage \cite{swipt1}.  Note that the total harvested  power by all users in the system ($P_H$) can be written as 
$
P_H= \sum_{i=1}^{C}\sum_{j=1}^{K_i}{P_{j,i}}.
$
\subsection{Problem Formulation}  
In this subsection, we formulate a problem to   investigate the EH capabilities of the   hybrid TDMA-NOMA scheme through a power minimization  (i.e., P-Min)  design. 
In particular, the BS aims to minimize the transmit power  (i.e., $P_t$) under minimum  harvesting power and   minimum data rate requirements at each user. Note that this design seeks to determine  a  power allocation  (i.e., $ p_{j,i}$)  and  a power splitting  ratio (i.e., $ \beta_{j,i}$) $ \forall i \in \mathcal{C}, \forall j \in \mathcal{K}_i$ for each user in the system. These design parameters can be determined through solving the following P-Min problem:
\begin{subequations}\label{min_power}
\begin{align}
 {OP_P\!:}~ 
 &\!\!\!\underset{\{p_{j,i},\beta_{j,i} \}_{i=1}^K}{\text{minimize }}\!\!\!
& &\!\!\!\!\!\!\sum_{i=1}^{C}  \sum_{j=1}^{K_i} p_{j,i}^2 \\
&  \text{subject to}
& &\!\!\!  {R}_{j,i}\geq  {R}^{\text{min}},  \forall i \in \mathcal{C}, \forall j \in \mathcal{K}_i,\label{min_power1}\\
      & &&\!\!\!   P_{j,i}\geq P^{\text{min}}, \forall i \in \mathcal{C}, \forall j \in \mathcal{K}_i, \label{min_power2}\\
      & &&\!\!\!   p_{K_i,i}^2\geq  \cdots\geq p_{1,i}^2, \forall i \in \mathcal{C},\label{min_power3}\\
      & &&\!\!\!    0\leq\beta_{j,i}\leq 1, \forall j \in \mathcal{K}_i, \forall i \in \mathcal{C}.\label{beta}
\end{align}
\end{subequations}
{ Note that ${R}^{\text{min}}$ and ${P}^{\text{min}}$ denote the minimum rate and minimum harvested power requirements at each user, respectively.}
It is obvious that   the optimization problem in (\ref{min_power}) is non-convex due to  the non-convex constraints in (\ref{min_power1}) and (\ref{min_power2}). Furthermore, the solution of  this optimization problem  needs to be determined by jointly optimizing the design parameters   $\{p_{j,i},\beta_{j,i} \}~ \forall i, \forall j$ . 

\section{Proposed Methodology and  Discussions}\label{sec3}
In this section, we propose an iterative approach to jointly solve  the original  non-convex optimization problem in (\ref{min_power}).  Later in this section, we  shed some light on the    initial parameters selection and convergence of the proposed iterative approach.  { We first provide a brief discussion on the proposed grouping strategy in the following subsection.}
 \subsection{Grouping Strategy}
   Choosing an appropriate grouping strategy plays a crucial role on the performance of the hybrid TDMA-NOMA system, as the optimal solution of $OP_P$ can be only determined by formulating the best groups \cite{cumanan}. In particular,  the optimal required transmit power  can be only determined by solving $OP_{p}$ with an exhaustive search among all possible set of groups \cite{faizeh2}. However, it is expensive in terms of computational complexity and  unaffordable in practical systems, including IoTs in future wireless networks.  Furthermore, the difference between the channel strengths of the users in the same group is another key factor that determines the overall performance of the system. As the users within each group are served based on NOMA, the practical implementation of the SIC technique requires the difference between the channel strengths of the users   to be as large as possible \cite{noma5g}.  Furthermore, grouping users with a similar channel strength would introduce  errors in the SIC implementation, which would degrade the overall performance of the system \cite{kim}. Based on this key fact, we group the users such that the difference between their channel strengths is as high as possible. For example, with two users in each group (i.e., $K_i=2$), the user groups based on the proposed grouping strategy can be defined as 
\begin{subequations}
\begin{align}
   \left(\{u_{1,1},u_{2,1}\}, \{u_{1,2},u_{2,2}\},\cdots \{u_{1,C},  u_{2,C}\}  \right)  \equiv \nonumber \\ \left(\{u_1,u_K\}, \{u_2,u_{K-1}\},\cdots \{u_{\frac{K}{2}},  u_{\frac{K}{2}+1}\}           \right). \label{group} 
   \end{align} 
\end{subequations}
   \subsection{Proposed Algorithm}
With the assumption that the users have been already grouped into groups as in (\ref{group}),  we now solve the non-convex optimization problem  $OP_P$  through the SCA technique. In SCA, the  non-convex terms are approximated by a set of lower bounded convex terms, and then, the original non-convex problem  is solved with these approximated convex problem. Note that SCA has been utilized
to solve different non-convex optimization problem in the literature   \cite{wcnc1}, \cite{wcnc2}, \cite{haitham}.   We start handling the non-convexity of the constraint in (\ref{min_power1}) by introducing new   slack variables $\vartheta_{j,i}$ and $\theta_{j,i}$,  such that, 
\begin{subequations}
\begin{align}
& {R}_{j,i} \geq \vartheta_{j,i}, \forall i \in \mathcal{C}, \forall j \in \mathcal{K}_i.\\
&  (1+\text{SINR}_{j,i}^d) \geq \theta_{j,i}, \forall i \in \mathcal{C}, \forall j \in \mathcal{K}_i, \forall d \in \{j+1, \cdots, \mathcal{K}_i\},\label{10n}\\
& \theta_{j,i} \geq 2^{\vartheta_{j,i}} \forall i \in \mathcal{C}, \forall j \in \mathcal{K}_i \label{100n}.
\end{align}
\end{subequations}
To handle the non-convexity of (\ref{10n}), we first insert a slack variable  $\chi_{j,i}$ such that
 \begin{multline}\label{1n}
\frac{\beta_{j,i}|h_{j,i}|^2p_{d,i}^2}{\beta_{j,i} |h_{j,i}|^2\sum_{s= 1}^{d-1} p_{s,i}^2+\beta_{j,i}\sigma_{j,i}^2+\tilde{\sigma}_{j,i}^2 }\geq \frac{(\theta_{j,i}-1) \chi_{j,i}^2}{\chi_{j,i}^2}, \\ \forall i \in \mathcal{C}, \forall j \in \mathcal{K}_i, \forall d \in \{j+1,j+2,\cdots, \mathcal{K}_i\}.
\end{multline} 
Then, we decompose the constraint in (\ref{1n}) into the following two constraints: 
\begin{multline}\label{2}
 {\beta_{j,i}|h_{j,i}|^2p_{d,i}^2}\geq (\theta_{j,i}-1) \chi_{j,i}^2, \\ \forall i \in \mathcal{C}, \forall j \in \mathcal{K}_i, \forall d \in \{j+1,j+2,\cdots, \mathcal{K}_i\},
\end{multline}
\begin{multline}\label{3n}
 \beta_{j,i} |h_{j,i}|^2\sum_{s= 1}^{d-1} p_{s,i}^2+\beta_{j,i}\sigma_{j,i}^2+\tilde{\sigma}_{j,i}^2 \leq \chi_{j,i}^2, \\ \forall i \in \mathcal{C}, \forall j \in \mathcal{K}_i, \forall d \in \{j+1,j+2,\cdots, \mathcal{K}_i\}.
\end{multline}
To handle the non-convexity of these constraints, we insert another  slack variable $\alpha_{j,i}^{d}$ such that
\begin{multline}\label{8n}
\beta_{j,i}  p_{d,i}^2\geq  \alpha_{j,i}^{d} \\ \forall i \in \mathcal{C}, \forall j \in \mathcal{K}_i, \forall d \in \{j+1,j+2,\cdots, \mathcal{K}_i\}.
\end{multline}
 It is obvious that the constraint in (\ref{8n}) is still non-convex. Therefore, we exploit the first-order Taylor series to approximate the left-hand side of (\ref{8n}) with a linear term. Doing so, the approximated convex form of (\ref{8n}) can be written as 
\begin{multline}\label{cona}
   \beta_{j,i}^{(t)}{p_{d,i}^2}^{(t)}+{2p_{d,i} }^{(t)}\beta_{j,i}^{(t)}\left({ p_{d,i} }-{p_{d,i} }^{(t)} \right)\\ +{p_{d,i}^2}^{(t)}\left(\beta_{j,i}-\beta_{j,i}^{(t)}\right) \geq \alpha_{j,i}^{d},\\ \forall i \in \mathcal{C}, \forall j \in \mathcal{K}_i, \forall d \in \{j+1,j+2,\cdots, \mathcal{K}_i\},
\end{multline}
where $\beta_{j,i}^{(t)}$ and ${p_{d,i}}^{(t)}$  represent the approximations of $\beta_{j,i} $ and ${p_{d,i}} $ at the $t^{th}$ iteration, respectively.
By incorporating  these  slack variables,  the non-convex constraint in (\ref{2}) can be approximately replaced by the following convex constraint:
 \begin{multline}\label{koko}
 |h_{j,i}|^2 \alpha_{j,i}^d \geq {\chi^2_{j,i}}^{(t)} \left(\theta_{j,i}^{(t)}-1 \right)+  2\left(\theta_{j,i}^{(t)}-1\right) {\chi_{j,i} }^{(t)}\\ \left(\chi_{j,i} -{\chi_{j,i} }^{(t)}\right)+ {\chi^2_{j,i} }^{(t)}\left(\theta_{j,i} -\theta_{j,i}^{(t)}\right)  , \\ \forall i \in \mathcal{C}, \forall j \in \mathcal{K}_i, \forall d \in \{j+1,j+2,\cdots, \mathcal{K}_i\}. 
  \end{multline}
 Note that the non-convex right-hand side of   inequality   (\ref{2}) is approximated with a linear term. Similarly, the approximated convex form of  constraint  (\ref{3n}) is written  as 
\begin{subequations}\label{soso1}
\begin{align}
 & {\chi^2_{j,i}}^{(t)}+2\chi_{j,i}^{(t)}\left(\chi_{j,i} -\chi_{j,i} ^{(t)}\right) \geq \nonumber \\& \gamma\left( |h_{j,i}|^2 \sum_{s=1}^{d-1} \alpha_{j,i}^s+\sigma_{j,i}^2 \beta_{j,i}+\tilde{\sigma}_{j,i}^2 \right). 
  \end{align}
  \end{subequations}
  Based on these multiple slack variables, the achieved rate at each user (i.e., $R_{j,i}$) can be equivalently approximated by   $ \vartheta_{j,i}$ with   the constraints in (\ref{100n}), (\ref{cona}), (\ref{koko}), and (\ref{soso1}).\\
 \indent Next, we  tackle the non-convexity of the constraint in (\ref{min_power2}) by  incorporating   slack variables  $\rho_{i,j}$ and $ \varrho_{i,j}$, such that
 \begin{subequations}
\begin{align}
& (1-\beta_{j,i}) p_{s,i}^2\geq \rho_{j,i}^s, \forall i \in \mathcal{C}, \forall j \in \mathcal{K}_i, \forall s \in\mathcal{K}_i,\label{jara}\\
 &\eta   |h_{j,i}|^2\sum_{s= 1}^{K_i} \rho_{j,i}^s   \geq   \varrho_{i,j}, \forall i \in \mathcal{C}, \forall j \in \mathcal{K}_i. 
\end{align}
\end{subequations} 
 Without loss of generality, the  non-convex constraint  in (\ref{jara}) can be tackled by following the same approximations  developed to handle the previous constraint in (\ref{8n}). Hence, 
  \begin{multline}\label{15}
  \left(1-\beta_{j,i}^{(t)}\right) {p_{s,i}^2}^{(t)}
  + 2 {p_{s,i} }^{(t)} \left(1-\beta_{j,i}^{(t)}\right)\left(p_{s,i}-p_{s,i} ^{(t)}\right)\\- {p_{s,i}^2}^{(t)}\left(\beta_{j,i}-\beta_{j,i}^{(t)}\right)  
  \geq \rho_{j,i}^s,   \forall i \in \mathcal{C}, \forall j \in \mathcal{K}_i, \forall s \in\mathcal{K}_i. 
  \end{multline}  
 {Finally, the non-convexity of the SIC constraint in (\ref{min_power3}) is tackled by  replacing each element in this constraint by the following linear approximation:
 \begin{equation}
p_{K_i,i}^2 \geq {p_{K_i,i}^2}^{(t)}+2p_{K_i,i}^{(t)}\left(p_{K_i,i}-p_{K_i,i}^{(t)}\right), \forall i. 
 \end{equation}}
  Based on these multiple slack variable incorporations, the non-convex optimization problem in (\ref{min_power}) can be equivalently written as 
\begin{subequations}\label{min_powerapp}
\begin{align}
 {{OP}_1\!:}~~ 
 &\!\!\!\underset{\Gamma} {\text{minimize }}\!\!\!
& &\!\!\!\!\!\!\sum_{i=1}^{C}  \sum_{j=1}^{K_i} p_{j,i}^2 \\
&  \text{subject to}
& &\!\!\! r_{j,i}\geq {R}^{\text{min}}, \varrho_{j,i}\geq P^{\text{min}}, \forall i \in \mathcal{C}, \forall j \in \mathcal{K}_i, \\
      & &&\!\!\!  (\ref{min_power3}), (\ref{beta}), (\ref{100n}), (\ref{cona}), (\ref{koko}),  (\ref{soso1}),(\ref{15}),
\end{align}
\end{subequations}
 where $ \Gamma$ includes all the optimization variables involved in the P-Min design: $ \Gamma={\{p_{j,i}, r_{j,i},\beta_{j,i},\varrho_{j,i}, \rho_{j,i}, \alpha_{j,i} \}_{i=1}^K}$. The solution of the  optimization problem in (\ref{min_powerapp}) requires an  appropriate  selection of the initial variables (i.e., $\Gamma^{(0)}$).  Therefore,  random  initial power allocations   ${\{p_{j,i}^{(0)} \}_{i=1}^K}$ are assumed. Then, the corresponding initial power splitting ratios (i.e., ${\{{\beta}_{j,i}^{(0)} \}_{i=1}^K})$ that satisfy the constraints of the original optimization problem $OP_P$, are evaluated. Furthermore, the remaining slack variables   $ \rho_{j,i}^{(0)}$  and $\alpha_{j,i}^{(0)}$   can be determined by substituting ${\{p_{j,i}^{(0)} \}_{i=1}^K}$ and  ${\{{\beta}_{j,i}^{(0)} \}_{i=1}^K}$ in (\ref{cona}) and (\ref{koko}), respectively. 
 The   algorithm proposed to solve the original P-Min problem is summarized in Algorithm 1. The algorithm terminates when the absolute difference between two sequential optimal values is less than a defined threshold $\mu$. 
 \begin{figure}[H]
    \hspace{0em}\hrulefill

 \hspace{0em} {\bf Algorithm 1:} P-Min Design using SCA.

     \hspace{0em}\hrulefill
     
     \hspace{0em} Step 1: Group  users based on (\ref{group})
     
     \hspace{0em} Step 2: Initialize all design parameters $\Gamma^{(0)}$

     \hspace{0em} Step 3: Repeat
 
\begin{enumerate}
   \item Solve the  optimization problem in (\ref{min_powerapp}) 
   \item Update $ \Gamma^{(n+1)} $ 
\end{enumerate}
\hspace{-0em} Step 4: Until required accuracy is achieved.

\hspace{-1em} \hrulefill
\end{figure} 

To demonstrate the performance of the proposed EH design  of the  hybrid TDMA-NOMA system, we  evaluate and compare its  performance with that of the conventional TDMA system.   
 In the TDMA system, each time slot ($t_{i}^{\text{TDMA}}=\frac{T}{K})$ is employed to serve only one user. Based on this time slot assignment, the achieved rate at $u_i$ can be written as 
\begin{equation}\label{tdma}
 {R}_i^{\text{TDMA}}=t_i^{\text{TDMA}} \log_2(1+\frac{\beta_i|h_i|^2p_i^2}{\beta_i\sigma_i^2+ \tilde{\sigma_i}^2}), \forall \in \mathcal{K}.
\end{equation}
 On the other hand,  the harvested power at $u_i$ in this conventional TDMA  can be represented as 
 \begin{equation}
 {P}_i^{\text{TDMA}}=\eta (1-\beta_i)|h_i|^2p_i^2, \forall \in \mathcal{K}.
 \end{equation} 
Now, we formulate a similar P-Min problem in a TDMA system with minimum rate and minimum energy harvesting constraints. As such, 
\begin{subequations}\label{dsdsd}
\begin{align}
 {OP_2\!:}~ ~
 &\!\!\!\underset{\{p_{i},\beta_{i} \}_{i=1}^K}{\text{minimize }}\!\!\!
& &\!\!\!\!\!\!\sum_{i=1}^{K}  p_{i}^2 \\
&  \text{subject to}
& &\!\!\!  {R}_{i}^{\text{TDMA}}\geq  {R}^{\text{min}},  \forall i \in \mathcal{K},  \label{min_psdsdowSDSer1}\\
      & &&\!\!\!   P_{i}^{\text{TDMA}}\geq P^{\text{min}}, \forall i \in \mathcal{K}. \label{min_posdsdweSDSr2}
\end{align}
\end{subequations}
Note that the developed optimization problem in (\ref{dsdsd}) for the TDMA system is solved using the same SCA technique.
 \vspace{-0.4in} 
 
\section{Simulation Results}\label{sec4}
\begin{figure}[!]
\centering
\includegraphics[width=1  \linewidth]{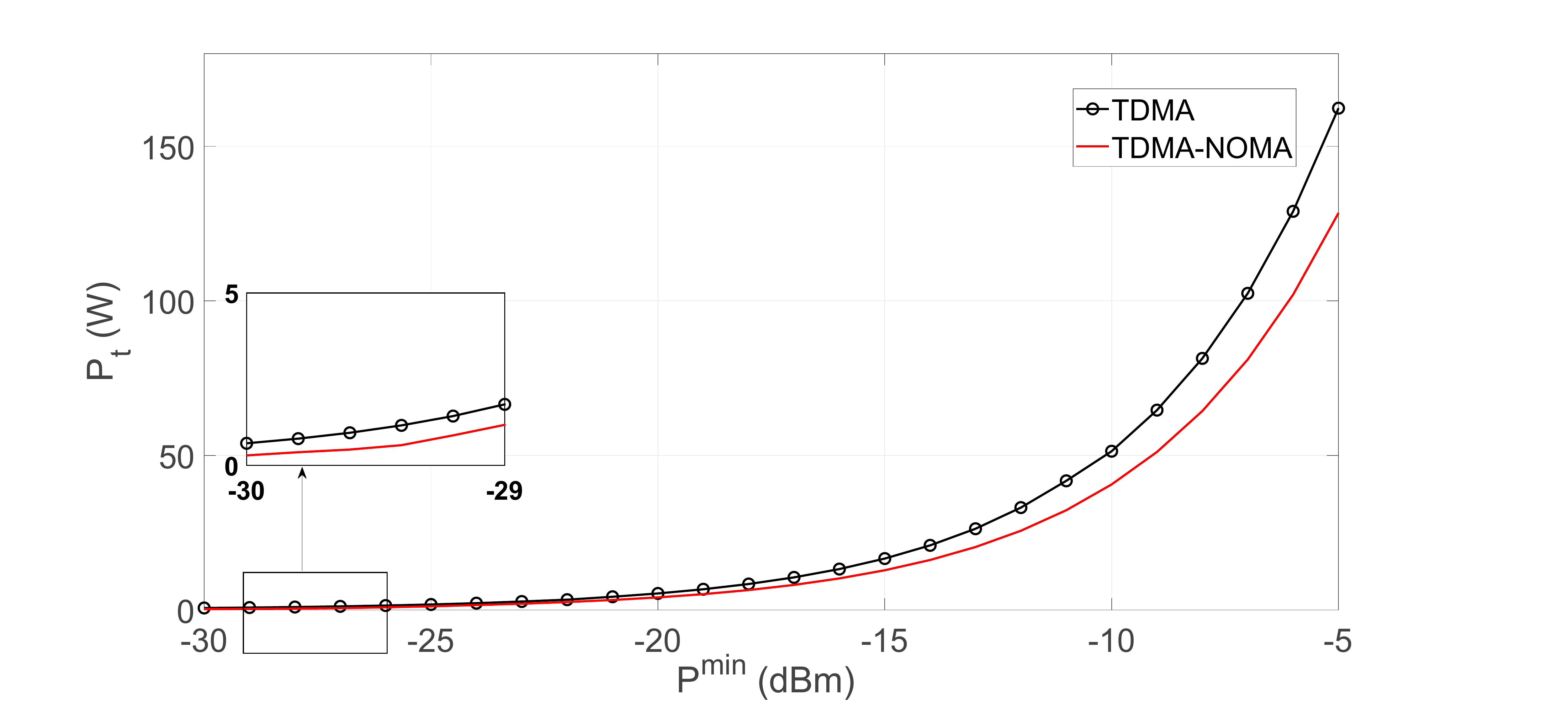}
\caption{The required transmit power versus different minimum harvest power requirements $P^\text{min}$,  with a minimum rate requirement $R^{\text{min}}= 10^{-1}$ bit/Hz.}
\label{figg}
\end{figure}

\begin{figure}[H]
\centering
\includegraphics[width=1 \linewidth]{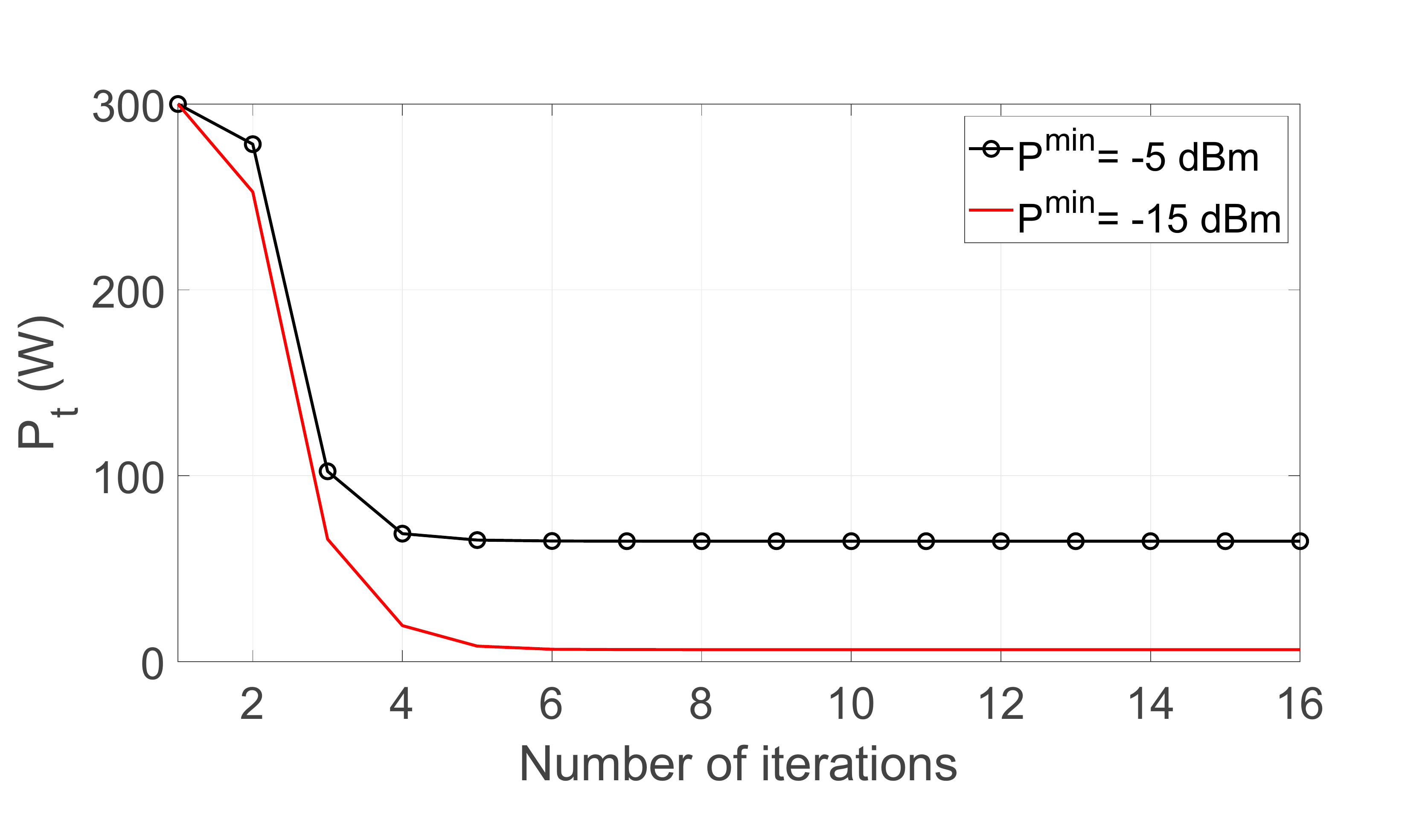}
\caption{The convergence of the SCA algorithm to solve $OP_P$ for different values of the minimum harvest power requirement  $P^\text{min}$, $R^{\text{min}}= 10^{-2}$ bit/Hz.}
\label{figg2}
\end{figure}

 \vspace{-0.2in} 
In this section, we demonstrate the EH capability of the proposed hybrid TDMA-NOMA scheme by evaluating and comparing its performance with that of  the conventional TDMA scheme. In these numerical simulations, we consider ten users (i.e., $K=10$) {  that are uniformly distributed in a circle of   radius 10 meter from the BS}. In addition, these users  are divided into five groups (i.e., $C=5$).  Table~\ref{tab:tab1} summarizes the  different parameters  adopted in   simulations \cite{swipt10}. {  Note that all simulation results are averaged over 500 channel realizations. In addition, the CVX toolbox   is used to generate    results in this section.}
\begin{table}[h]
\caption{Parameter values used in   simulations.}
\centering
\label{tab:tab1}
 {\begin{tabular}{|c|c|}
\hline

{\bf Parameter} & {\bf Value}\\
\hline
Number of users ($K$)  & $10$ \\
\hline
Number of groups ($C$) & $5$\\
\hline
Number of users in each group ($K_i$)& 2\\
 
\hline
  Path loss exponent ($ \kappa$)  & $2$\\
  \hline
  Reference distance ($ d_0$)  & $1$\\
  \hline
  Signal attenuation at $ d_0$ ($ \eta$)  & $-30$ dB\\
  \hline
$\sigma_i^2$,  $\hat{\sigma}_{j,i}^2$, $\tilde{\sigma}_{j,i}^2$ ( dBm/Hz)& $-100$ \\
\hline
 
  Efficiency of converter  ($ \eta)$  &  $0.75$\\
\hline
 
\end{tabular}}
\end{table}
Fig. \ref{figg} illustrates and compares the minimum required transmit  power (i.e., P$_t$) against a range of   minimum harvest power requirements P$^\text{min}$ for the    hybrid TDMA-NOMA and the conventional TDMA systems.  As expected,   P$_t$ of both systems increases with the increase of P$^\text{min}$. Furthermore, as shown in Fig. \ref{figg}, the hybrid TDMA-NOMA scheme exhibits a better performance, as it consumes less P$_t$ compared to the conventional TDMA system.  In particular,  users grouping in the hybrid TDMA-NOMA system   introduces     higher interference levels,   which facilitates  the satisfaction of the minimum harvested power requirements with lower $P_t$ compared to the conventional TDMA system.
Next, we evaluate the number of iterations required for the convergence of SCA algorithm to solve $OP_P$ for two different minimum harvested power requirements.   As seen in 
   Fig. \ref{figg2},   the algorithm converges to the solution within a few number of iterations.

\section{Conclusion}\label{sec5}
In this paper, we investigated the energy  harvesting   capabilities of a multi-user SISO hybrid TDMA-NOMA system. In such a hybrid system,   users are divided into a number of  groups,  with a time slot    assigned to serve each group and NOMA employed to serve users within a group. For the proposed scheme, we evaluated the required minimum power   to meet   the minimum rate  and minimum harvest energy requirements at each user.    { Simulation results confirmed that the  proposed hybrid TDMA-NOMA system outperforms the conventional TDMA system in terms of the  minimum required transmit power.}

  {  \section*{Acknowledgement}
The work of H. Al-Obiedollah was supported by The Hashemite University, Jordan.  The work of K. Cumanan, A.  Burr and Z. Ding was supported by H2020-MSCA-RISE-2015 under grant no: 690750. The work of Y. Rahulamathavan was supported by UK-India Education Research Initiative (UKIERI) through grant UGC-UKIERI-2016-17-019.  

}
 
\bibliography{references}
  \bibliographystyle{ieeetran}

\end{document}